\documentclass[aps,prl,twocolumn,showpacs,superscriptaddress]{revtex4}
\usepackage{graphicx}

\begin{document}

\title{ Effect of Li-deficiency impurities on the electron-overdoped LiFeAs
    superconductor}
\author{Meng Wang}
\affiliation{Beijing National Laboratory for Condensed Matter Physics, Institute of Physics, Chinese Academy of Sciences, Beijing
100190, China }
\author{Miaoyin Wang}
\affiliation{Department of Physics and Astronomy, The University of Tennessee, Knoxville,
Tennessee 37996-1200, USA }
\author{Hu Miao}
\affiliation{Beijing National Laboratory for Condensed Matter Physics, Institute of Physics, Chinese Academy of Sciences, Beijing
100190, China }
\author{S. V. Carr}
\affiliation{Department of Physics and Astronomy, The University of Tennessee, Knoxville,
Tennessee 37996-1200, USA }
\author{D. L. Abernathy}
\affiliation{Quantum Condensed Matter Division, Oak Ridge National Laboratory, Oak
Ridge, Tennessee 37831-6393, USA}
\author{M. B. Stone}
\affiliation{Quantum Condensed Matter Division, Oak Ridge National Laboratory, Oak
Ridge, Tennessee 37831-6393, USA}
\author{X. C. Wang}
\affiliation{Beijing National Laboratory for Condensed Matter Physics, Institute of Physics, Chinese Academy of Sciences, Beijing
100190, China }
\author{Lingyi Xing}
\affiliation{Beijing National Laboratory for Condensed Matter Physics, Institute of Physics, Chinese Academy of Sciences, Beijing
100190, China }
\author{C. Q. Jin}
\affiliation{Beijing National Laboratory for Condensed Matter Physics, Institute of Physics, Chinese Academy of Sciences, Beijing
100190, China }
\author{Xiaotian Zhang}
\affiliation{Beijing National Laboratory for Condensed Matter Physics, Institute of Physics, Chinese Academy of Sciences, Beijing
100190, China }
\author{Jiangping Hu}
\affiliation{Beijing National Laboratory for Condensed Matter Physics, Institute of Physics, Chinese Academy of Sciences, Beijing
100190, China }
\affiliation{Department of Physics, Purdue University, West Lafayette, Indiana 47907, USA}
\author{Tao Xiang}
\affiliation{Beijing National Laboratory for Condensed Matter Physics, Institute of Physics, Chinese Academy of Sciences, Beijing
100190, China }
\author{Hong Ding}
\affiliation{Beijing National Laboratory for Condensed Matter Physics, Institute of Physics, Chinese Academy of Sciences, Beijing
100190, China }
\author{Pengcheng Dai}
\email{pdai@utk.edu}
\affiliation{Department of Physics and Astronomy, The University of Tennessee, Knoxville,
Tennessee 37996-1200, USA }
\affiliation{Beijing National Laboratory for Condensed Matter Physics, Institute of Physics, Chinese Academy of Sciences, Beijing
100190, China }

\begin{abstract}
We use transport, inelastic neutron scattering, and angle resolved photoemission experiments to demonstrate
that the stoichiometric LiFeAs is an intrinsically electron-overdoped superconductor
similar to those of the electron-overdoped NaFe$_{1-x}T_x$As and BaFe$_{2-x}$$T_x$As$_2$ ($T=$ Co,Ni).
Furthermore, we show that although transport properties of the stoichiometric superconducting LiFeAs and Li-deficient
nonsuperconducting Li$_{1-x}$FeAs are different, their
electronic and magnetic properties are rather similar.  Therefore, the nonsuperconducting Li$_{1-x}$FeAs is also
in the electron overdoped regime, where small Li deficiencies near the FeAs octahedra can dramatically suppress superconductivity
through the impurity scattering effect.
\end{abstract}

\maketitle




Superconductivity in iron pnictides occurs near an antiferromagnetic (AF) instability \cite{kamihara,cruz}.  When the AF order in a nonsuperconducting (NSC) parent compound is suppressed by electron or hole doping, superconductivity emerges with
the persistent short-range spin excitations directly coupled to the superconducting (SC) transition temperature $T_c$ \cite{lumsden,lynn}.
While this general behavior is obeyed in most iron pnictide superconductors and suggests the importance of magnetism to the superconductivity of these materials \cite{lumsden,lynn},
the only exception is the stoichiometric LiFeAs (Fig. 1a), which does not have a static AF ordered parent compound and superconducts with a relatively high $T_c$ of $\sim$17 K without any doping \cite{xcwang,jhtapp,mjpitcher,flpratt,cwchu}.  Furthermore, a few percent of Li deficiencies in Li$_{1-x}$FeAs can increase the resistivity and
destroy superconductivity  (Figs. 1b and 1c) \cite{xcwang}.  Indeed, the absence of the AF order in LiFeAs is caused by the poor nesting condition between the
shallow hole-like Fermi pocket near the
$\Gamma (0,0)$ point and the large electron Fermi surface at the $M(1,0)/(0,1)$
points in the Brillouin zone (Fig. 1d) \cite{borisenko}.  These observations have fueled the suggestion that
the mechanism of superconductivity in LiFeAs is due to ferromagnetic instability and $p$-wave triplet pairing
\cite{borisenko,brydon,hanke}.  This is
fundamentally different from all other iron pnictides, where the singlet electron pairing superconductivity and AF order are both believed
to be associated with the sign reversed quasiparticle excitations between the hole and electron Fermi surfaces near the $\Gamma (0,0)$ and $M(1,0)/(0,1)$
points \cite{Hirschfeld,kuroki,chubukov,fwang}.

Here we describe transport, inelastic neutron scattering, and angle resolved photoemission (ARPES)
experiments on the stoichiometric SC LiFeAs and Li-deficient NSC Li$_{1-x}$FeAs.
We find that a few percent Li-deficiencies in Li$_{1-x}$FeAs can completely suppress
superconductivity and change transport properties but
without much affecting the sizes of the Fermi surfaces \cite{umezawa} and incommensurate spin excitations in the SC LiFeAs \cite{qureshi}.
By comparing our
results with previous work on the SC LiFeAs \cite{umezawa,qureshi,aetaylor},
NaFe$_{1-x}$Co$_x$As \cite{slli,drparker,xdzhou}, BaFe$_{2-x}T_x$As$_2$ ($T=$ Co, Ni) \cite{clester10,jtpark10,huiqian},
and LaFe$_{1-y}$Zn$_y$AsO$_{1-x}$F$_x$ \cite{ykli}, 
we conclude that the stoichiometric LiFeAs is an intrinsically electron-overdoped superconductor, and that Li deficiencies affect its SC properties similar to the Zn impurity effects in the electron-overdoped iron
pnictide superconductors \cite{ykli}.  Therefore, the mechanism of superconductivity in LiFeAs is associated with
AF spin excitations and not fundamentally different from all other
iron-based superconductors \cite{Hirschfeld,kuroki,chubukov,fwang}.

\begin{figure}[t]
\includegraphics[scale=0.4]{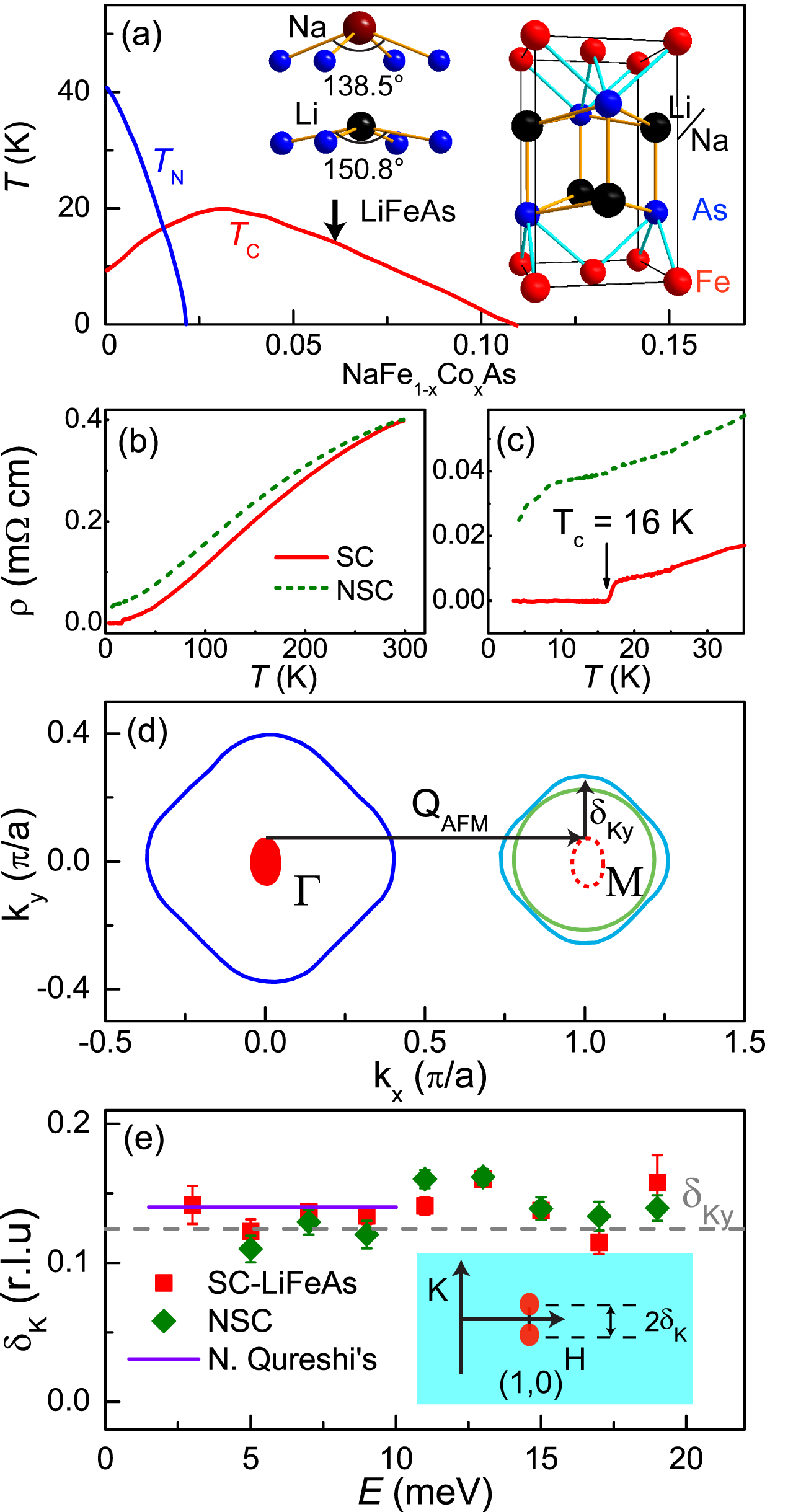}
\caption{ (color online).
(a) Phase diagram of the electron doped NaFe$_{1-x}$Co$_x$As
from Ref. \cite{drparker}. The inset shows the structure of Na(Li)FeAs and
  the differences in the angles of the two alkali arsenic planes
based on the structural parameters from Ref. \cite{mjpitcher} for LiFeAs and Ref. \cite{slli} for NaFeAs.
(b) The temperature dependence of resistivity for the SC LiFeAs (solid line) and NSC Li$_{0.94}$FeAs (dashed line) up to room temperature.
The data are normalized by the size and mass of the single crystals.
(c) Expanded view of the temperature dependence of the resistivity for the SC and NSC Li$_{1-x}$FeAs.
 The SC LiFeAs has a clear transition to superconductivity at 16 K.
(d) Schematic Fermi surfaces of LiFeAs from Ref. \cite{umezawa}. The red shadow indicates a flat band in the center of the $\Gamma(0,0)$ point.
The incommensurability from the ARPES measurements is defined as $\delta_K$, the mismatch of the inner hole Fermi surface and electron Fermi surfaces.
(e) The energy dependence of the incommensurability for the incommensurate
spin excitations from the SC LiFeAs (the red squared symbols), NSC Li$_{0.94}$FeAs (the olive diamond symbols), and the APRES measurements (the grey dash line).
The violet solid line is the incommensurability value from Ref. \cite{qureshi}.
The inset shows the locations of the incommensurate peaks near the in-plane AF wave vector ${\bf Q}=(1, 0)$ in LiFeAs.
}
\end{figure}

Our transport measurements on the SC and NSC Li$_{1-x}$FeAs
 were carried out on a commercial physical properties measurement system using the
 four probe method.  The inelastic neutron scattering experiments were performed on
the ARCS time-of-flight chopper
spectrometer at the spallation neutron source, Oak Ridge National
laboratory \cite{mwang11}. The ARPES experiments were performed at the PGM beamline of the Synchrotron Radiation Center,
Stoughton, Wisconson. The energy and angular resolutions
of the ARPES measurements
were set at $\pm 20$ meV and 0.2$^\circ$, respectively. The samples were cleaved in situ and measured at 20 K in a vacuum better than $4\times10^{-11}$ Torr.
The incident photon energy was chosen to be $h\nu=35$ eV.
Our single crystals of the SC LiFeAs were grown using the self flux method with the
 $^7$Li isotope to minimize the neutron absorption effect.
The method for growing the NSC Li$_{1-x}$FeAs with natural Li were described previously \cite{mwang11}.
 The inductively coupled plasma analysis on the samples showed that the
compositions of the NSC crystals are Li$_{0.94\pm 0.01}$FeAs \cite{mwang11}.  To within the errors of our measurements,
the SC LiFeAs was found to be stoichiometric.
Figure 1b shows temperature dependence of the
resistivity for the SC and NSC Li$_{1-x}$FeAs. Figure 1c plots the expanded view of the low-temperature
resistivity for both samples, which reveals a $T_c=16$ K for the SC LiFeAs and larger resistivity
for the NSC Li$_{0.94}$FeAs.  For inelastic neutron scattering measurements on ARCS, we co-aligned approximately 
3.95 grams of the SC single crystals of LiFeAs with a mosaic of 2$^{\circ }$.  The NSC Li$_{0.94}$FeAs was the same sample used
in our previous measurements \cite{mwang11}.  They are 
mounted inside a He exchange gas filled 
thin aluminum can which was mounted directly to the cold-finger of a closed cycle
 He refrigerator for ARCS measurements,
where the wave vector $Q$ at ($q_{x}$, $q_{y}$, $q_{z}$) in \AA$^{-1}$ is defined as
$(H,K,L)=(q_{x}a/2\pi ,q_{y}b/2\pi ,q_{z}c/2\pi )$ reciprocal lattice units
(rlu) with $a=b=5.316$ \AA , and $c=6.306$ \AA.

\begin{figure}[t]
\includegraphics[scale=0.5]{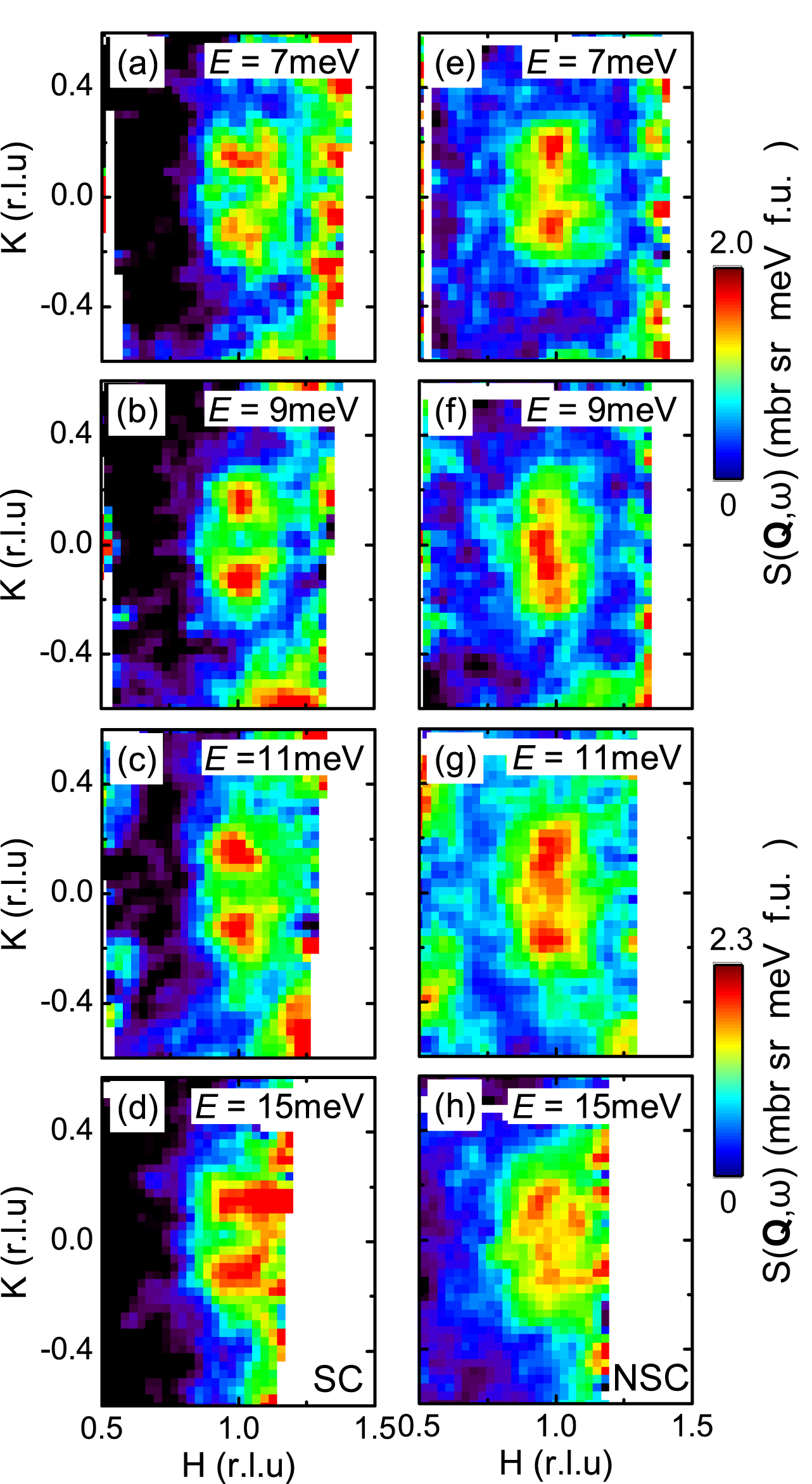}
\caption{ (color online).
Two-dimensional constant-energy images of the spin excitations in the $[H, K]$ plane
at the spin excitation energies indicated (a-d) for the SC LiFeAs and (e-h) for the NSC Li$_{0.94}$FeAs at 5 K.
The incident beam neutron energy was chosen to be
$E_i = 35$ meV along the $c$-axis.  The intensity has been normalized to be in absolute units using a vanadium standard as
discussed earlier \cite{mwang11}.
}
\end{figure}

In our previous inelastic neutron scattering work on the NSC Li$_{0.94}$FeAs with natural Li \cite{mwang11},
we have reported the presence of a large spin gap of $\Delta =13$ meV
at the AF ordering wave vector $Q=(1,0,3)$
 using triple-axis spectroscopy.  The gap was found to be
temperature independent between 2 and 190 K \cite{mwang11}.  More recently, inelastic neutron scattering
experiments on the SC LiFeAs with the
 $^7$Li isotope found low-energy ($1.5\le E\le 13$ meV)  transverse
 incommensurate spin excitations that appear to couple to $T_c$ \cite{qureshi}. In the light of this development, we have carried out new measurements on ARCS with the incident neutron beam
direction parallel to the $c$-axis and
 $E_{i}=35$ meV for both the SC LiFeAs and NSC Li$_{0.94}$FeAs at 5 K.   For the NSC Li$_{0.94}$FeAs with highly neutron absorbing $^6$Li, the scattering
 geometry at ARCS is much better than the earlier triple-axis measurements \cite{mwang11} because neutrons only have to pass the thinest part of the platelet samples.
 Figure 2
summarizes the outcome of these measurements.  For the SC LiFeAs, Figures 2a-2d show the
two-dimensional constant-energy ($E$)
images of the scattering in the $(H,K)$ plane for $E=7\pm 1$,
$9\pm 1$, $11\pm 1$, and $15\pm 1$ meV,  respectively.  Consistent with previous work \cite{qureshi}, we can see clear transverse incommensurate peaks centered near the in-plane AF wave vector ${\bf Q}=(1,0)$ at all the probed energies.  Figures 2e-2h plot two-dimensional scattering images for the NSC Li$_{0.94}$FeAs
at $E=7\pm 1$, $9\pm 1$, $11\pm 1$, and $15\pm 1$ meV,  respectively.  These results reveal the presence of low-energy spin excitations in the NSC Li$_{0.94}$FeAs, different from the earlier triple-axis measurement \cite{mwang11}.
While spin excitations are clearly incommensurate at the probed energies for the SC LiFeAs (Figs. 2a-2d), the
incommensurability is less well-defined for the NSC Li$_{0.94}$FeAs (Figs. 2e-2h).

\begin{figure}[t]
\includegraphics[scale=0.55]{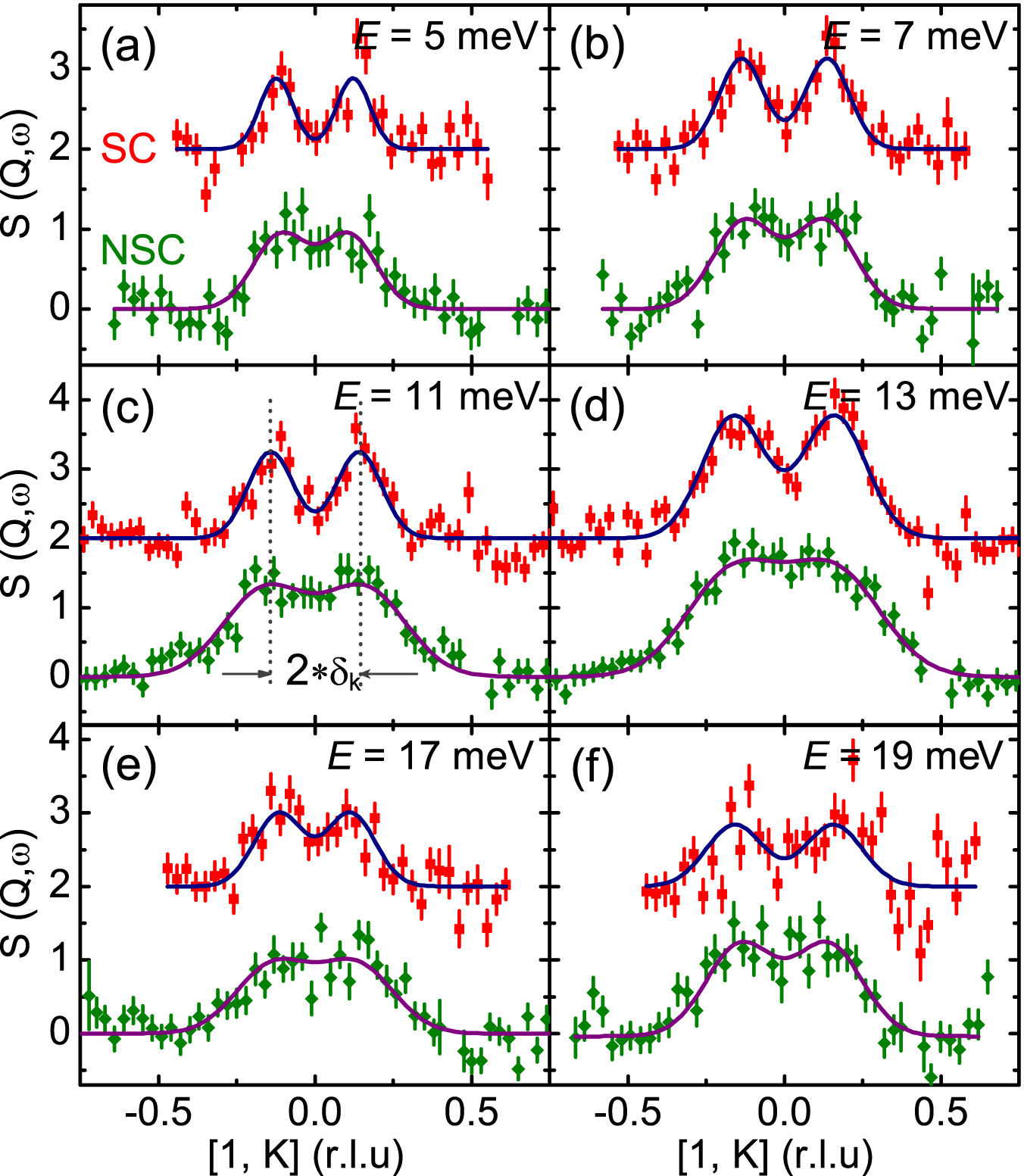}
\caption{(color online). Constant-energy cuts of spin excitations
along the $[1,K]$ direction for the SC LiFeAs and NSC
Li$_{0.94}$FeAs at energy
transfers of (a) $E=5\pm 1$ meV; (b) $7\pm 1$ meV;
(c) $11\pm 1$ mV; (d) $13\pm 1$ meV; (e) $17\pm 1$ meV; (f) $19\pm 1$
meV; all with $E_{i}=35$ meV. The solid lines are fits to two Gaussian peaks. The dashed vertical lines in (c), marked the center of peaks, indicate the definition of incommensurability of spin excitations as in previous work \cite{qureshi}. The cuts for the SC and NSC spin excitations spectra
were subtracted by the same fitted NSC background at the identical energy. The intensity is in absolute units, and error bars indicate one standard deviation.
For presentation, the SC data are offset vertically by 2 units for all panels.
 }
\end{figure}

\begin{figure}[t]
\includegraphics[scale=0.4]{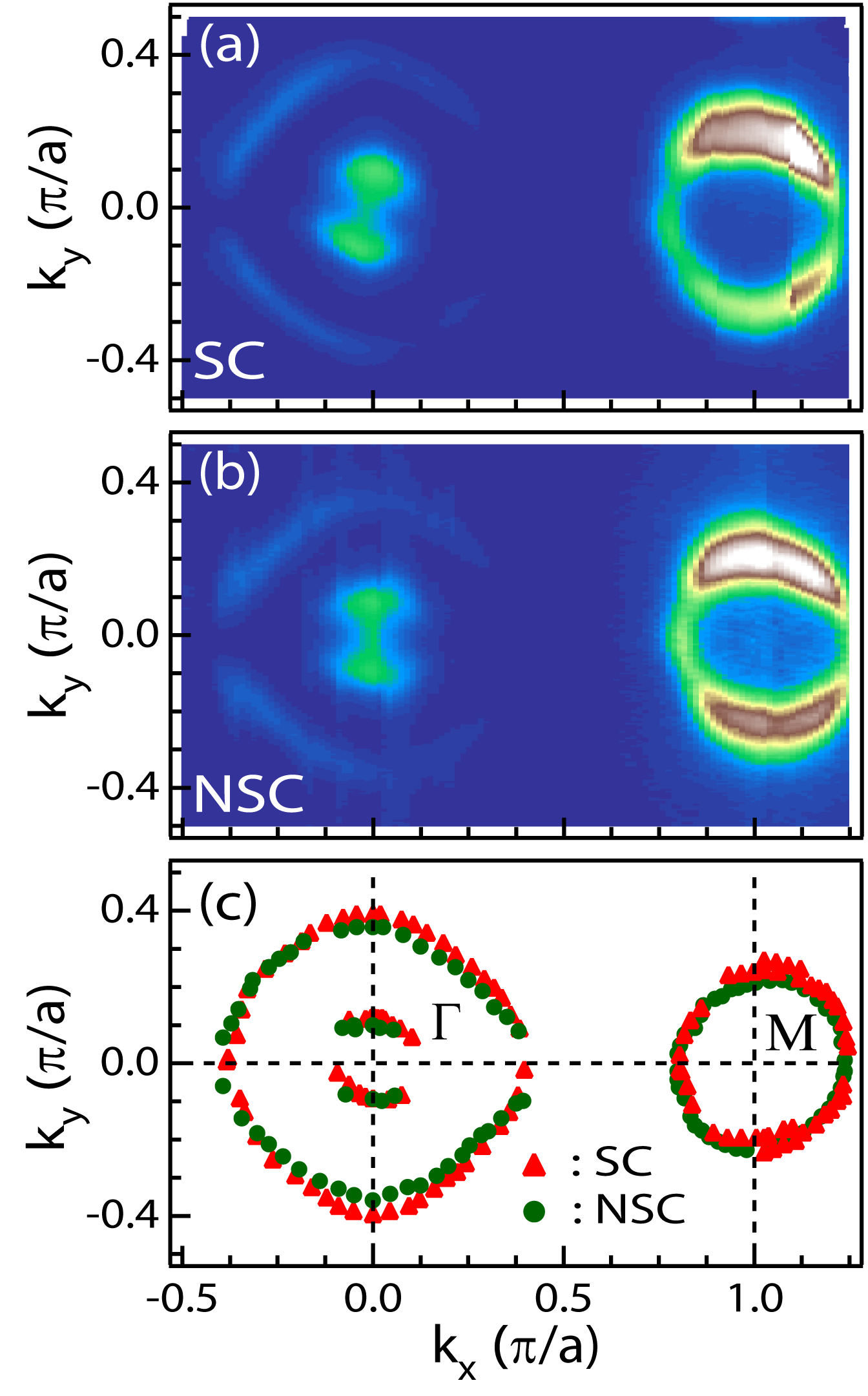}
\caption{(color online).
(a) and (b) ARPES intensity mappings of the SC and NSC Li$_{1-x}$FeAs samples recorded with $h\nu = 35$ eV photons
(Corresponding to the $c$-axis momentum transfer $k_Z =\pi$) and integrated within  $\pm 20$ meV with respect to the
Fermi energy $E_F$. The extracted Fermi surface contours from (a) and (b) are plotted together in (c).
 }
\end{figure}

To quantitatively determine the differences in spin excitations of the SC and NSC Li$_{1-x}$FeAs, we
cut through the transverse direction of the
two-dimensional scattering images in Fig. 2 for both samples.
 Figures 3a-3f show constant-energy cuts along
the $[1,K]$ direction for energies of $E=5\pm 1$, $7\pm 1$, $11\pm 1$,
$13\pm 1$, $17\pm 1$, and $19\pm 1$ meV, respectively.
Inspection of the Figure reveals that the incommensurabilities of spin excitations for both the SC and NSC Li$_{1-x}$FeAs are very similar and nearly 
energy independent within the probed excitation energy range of $5\le E\le 19$ meV.  However, the incommensurate spin excitations
in the SC LiFeAs have better defined peaks with
longer spin-spin correlation lengths compared with that of the NSC Li$_{0.94}$FeAs (Fig. 3).  
Simple Gaussian fits to the data in Fig. 3 are able to extract the incommensurate peak position, $\delta_K$, as a function of energy transfer.  This is shown in Fig. 1e, again illustrating the similar amount of and the lack of change of incommensurability with energy transfer in both compounds.
Based on these data, we see that the low-energy
spin excitations in the superconductivity-suppressed Li$_{0.94}$FeAs are remarkable similar to those of the
SC LiFeAs.  Therefore, Li-deficiency induced suppression of superconductivity does not fundamentally alter the magnetic properties of the
 SC LiFeAs.

If we assume that the Li-deficiencies in Li$_{1-x}$FeAs remove electrons from the FeAs octahedra, the SC LiFeAs should have a larger electron-doping level than that of the
NSC Li$_{0.94}$FeAs and therefore should have a larger electron
Fermi surface size. Figures 4a and 4b show the ARPES intensity mappings of the SC LiFeAs and NSC  Li$_{0.94}$FeAs, respectively.
Figure 4c plots the corresponding hole and electron Fermi pockets near the $\Gamma (0,0)$ and $M(1,0)/(0,1)$
points, respectively, for the SC and NSC samples.  To within the errors of our measurements, we find that the SC
LiFeAs and NSC Li$_{0.94}$FeAs have the same Fermi surface topology (Fig. 4c).  Therefore, a few percent Li-deficiencies
in Li$_{1-x}$FeAs do not dramatically change the hole and electron Fermi pocket sizes and alter the Fermi surface nesting conditions.  This is consistent with
the similar incommensurate spin excitations in the SC and NSC Li$_{1-x}$FeAs (Figs. 2 and 3), but contrary to the naive expectation that
the Li-deficiencies in Li$_{1-x}$FeAs should reduce the sizes of the electron Fermi surface and enlarge the hole Fermi surface.

In previous work, nonmagnetic
Zn impurities were found to severely suppress superconductivity for
LaFeAsO$_{1-x}$F$_x$ in the electron-overdoped regime but were much less effective in reducing $T_c$ for the under and optimally electron doped samples \cite{ykli}.
Similarly, Zn impurities were found to be effective in suppressing superconductivity in BaFe$_{2-x}$Co$_x$As$_2$ \cite{junli}.
 This behavior is consistent with the $s^{\pm}$-wave SC state, where the nonmagnetic impurity scattering should rapidly decrease $T_c$ \cite{rmfernandez}.  If we assume
that the Li-vacancies in Li$_{1-x}$FeAs indeed have a limited impact on the size of the electron and hole Fermi surfaces,
the rapid suppression of superconductivity by small amount of Li-deficiencies may indicate that superconductivity in the stoichiometric LiFeAs is in the electron-overdoped
regime \cite{rmfernandez}.
To see why this may be the case, we consider the lattice structure of LiFeAs  as shown in the inset of Fig. 1a.  Comparing with
the AF ordered Na$_{1-\delta}$FeAs \cite{slli}, the FeAs octahedron in LiFeAs is much more compressed with an Fe-As distance of $\sim$2.417 \AA\ \cite{mjpitcher} similar to the Fe-As distance of $\sim$2.42 \AA\ 
in the electron-overdoped NaFe$_{1-x}$Co$_x$As with $x=0.2$ \cite{drparker}.  To understand why small Li-deficiencies can dramatically suppress superconductivity
of the stoichiometric LiFeAs while the Na-deficiencies in Na$_{1-\delta}$FeAs actually promotes superconductivity \cite{tanatar},
we note that the Na ions in Na$_{1-\delta}$FeAs form a buffer layer rather far removed from the FeAs octahedra whereas the Li ions in LiFeAs are almost in the
As-layer of the FeAs octahedra \cite{mjpitcher,drparker}.  

Assuming that the electric conductivity in LiFeAs arises from the hopping of itinerant electrons between
the Fe atoms through the As bridge, the Li-vacancies in LiFeAs near the FeAs octahedra can act as impurity centers which scatter off
conduction band electrons.  The correction to $T_c$ by the impurity scattering is a universal function of
the impurity scattering rate $\Gamma$.  For the $s^{+-}$-wave superconductor, it
was shown that the SC
transition temperature is completely suppressed if the ratio between $\Gamma$ and the $T_c$ value without impurities
is approximately larger than 1 \cite{Bang09}.
The value of $\Gamma$ can be estimated from the resistivity difference $\Delta\rho_i$ between the Li-deficient
and stoichiometric Li$_{1-x}$FeAs via $\Delta\rho_i = m^\ast \Gamma/ e^2 n$, where $m^\ast$ is the
effective mass of quasiparticle and $n$ is the electron density per unit cell.
From Figure 1c, we see that $\Delta\rho_i$ is about 0.03 $\mathrm{m\Omega\, cm}$.
For LiFeAs, the effective mass is $\sim$5 times the 
bare electron mass \cite{putzke}.  If there are two itinerant electrons per Fe, 
 we find that $\Gamma \approx 2.2 T_{c}$, which is
larger than the critical value of $\Gamma$ that is needed for completely suppressing $T_c$. This is
consistent with the picture that 
the out-of-plane Li-vacancies in LiFeAs play the same role as the nonmagnetic Zn impurities in
the electron-overdoped LaFe$_{1-y}$Zn$_y$AsO$_{1-x}$F$_x$ \cite{ykli}.

If the stoichiometric LiFeAs is indeed an electron-overdoped superconductor, there should be no static AF order in LiFeAs as that in NaFeAs \cite{slli}.  Instead, the quasiparticle excitations between the mismatched hole and electron Fermi surfaces due to the self electron-doping should produce incommensurate spin fluctuations along the direction transverse to the AF ordering wave vector ${\bf Q}=(1,0)$ (Figs. 1d and 1e) consistent with the calculated 
spin susceptibility $\chi^{\prime\prime}({\bf Q},\omega)$ 
based on a random phase approximation of a three-dimensional 5-orbital tight-binding model for BaFe$_2$As$_2$ \cite{jtpark10,huiqian,graser}.   
Experimentally, the
transverse incommensurate spin fluctuations with $\delta_K\approx 0.1$
were found at $E=7$ meV for the electron overdoped BaFe$_{2-x}$Ni$_x$As$_2$ at $x=0.15$ \cite{huiqian}.  Using the ARPES measurements (Fig. 4), we plot
in Fig. 1d the hole and electron Fermi surfaces of the SC and NSC
Li$_{1-x}$FeAs.  Assuming that the large hole pocket near $\Gamma (0,0)$ is unfavorable for the Fermi surface nesting, we see that
the nesting of the small hole pocket near $\Gamma (0,0)$ and the electron pockets near $M(1,0)/(0,1)$ should give
incommensurate spin excitations at $\delta_{K}$ as shown in Fig. 1e.  This nesting condition is consistent with
previous work on LiFeAs \cite{qureshi} and our own measurements.  These results are also in agreement with the Fermi surface nesting interpretation of
the low-energy spin excitations in
the electron- \cite{jtpark10,huiqian} and hole-doped \cite{clzhang,castellan} BaFe$_2$As$_2$, and thus suggesting that the stoichiometric LiFeAs is an intrinsically electron-overdoped superconductor.

Moreover, recent systematic scanning tunneling microscopy (STM) measurements on NaFe$_{1-x}$Co$_x$As reveal that the
tunneling spectra $dI/dV$ change from the symmetric lineshape around Fermi energy for the optimally electron doped sample ($x=0.028$) to
a strong asymmetric lineshape in the electron overdoped regime ($x=0.061$) \cite{xdzhou}.
Since STM measurements on the SC LiFeAs (see Fig. 1b in
Ref. \cite{hanke}) show strong asymmetric tunneling spectra consistent with that of the electron overdoped NaFe$_{1-x}$Co$_x$As with $x=0.061$ \cite{xdzhou},
it is inevitable that the SC and NSC Li$_{1-x}$FeAs
are in the electron overdoped regime similar to the electron overdoped NaFe$_{1-x}$Co$_x$As (Fig. 1a).

We thank P. Richard, T. Qian, Zhuan Xu, Shiliang Li, and Yayu Wang for helpful discussions.
The work in IOP is supported by the MOST of China through 973 projects: 2012CB821400,
2010CB833102, and 1J2010CB923001. The work at UTK
is supported by the U.S. DOE BES No. DE-FG02-05ER46202.
The research at ORNL's SNS was sponsored by the Scientific User Facilities Division, BES, U.S. DOE.
The ARPES work at SRC is primarily funded by the University of Wisconsin-Madison with supplemental support from facility Users and the University of Wisconsin-Milwaukee.


\begin{thebibliography}{99}

\bibitem{kamihara} Y. Kamihara {\it et al.}, J.
Am. Chem. Soc. \textbf{130}, 3296 (2008).

\bibitem{cruz} C. de la Cruz \textit{et al.}, Nature (London) \textbf{453},
899 (2008).

\bibitem{lumsden} M. D. Lumsden and A. D. Christianson, J. Phys.: Condens.
Matter \textbf{22}, 203203 (2010).

\bibitem{lynn} J. W. Lynn and P. C. Dai, Physica C \textbf{469}, 469 (2009).

\bibitem{xcwang} X. C. Wang \textit{et al.}, Solid State Commun. \textbf{148}
, 538 (2008).

\bibitem{jhtapp} J. H. Tapp \textit{et al.}, Phys. Rev. B \textbf{78},
060505(R) (2008).

\bibitem{mjpitcher} M. J. Pitcher \textit{et al.}, Chem. Commun., 5918
(2008).

\bibitem{flpratt} F. L. Pratt \textit{et al.}, Phys. Rev. B \textbf{79},
052508 (2009).

\bibitem{cwchu} C. W. Chu \textit{et al.}, Physica C \textbf{469}, 326
(2009).

\bibitem{borisenko} S. V. Borisenko \textit{et al.}, Phy. Rev. Lett. \textbf{
105}, 067002 (2010).

\bibitem{brydon} P. M. R. Brydon {\it et al.}, Phys. Rev. B \textbf{83}, 060501(R) (2011).

\bibitem{hanke} T. H$\rm \ddot{a}$nke {\it et al.}, Phys. Rev. Lett. {\bf 108}, 127001 (2012).

\bibitem{Hirschfeld} P. J. Hirschfeld, M. M. Korshunov, I. I. Mazin, Rep. Prog. Phys. {\bf 74}, 124508 (2011).

\bibitem{kuroki} K. Kuroki {\it et al.}, Phys. Rev. Lett. \textbf{101}, 087004
(2008).

\bibitem{chubukov} A. V. Chubukov, Annu. Rev. Condens. Matter Phys. {\bf 3}, 13 (2012).

\bibitem{fwang} F. Wang \textit{et al.}, Phys. Rev. Lett. \textbf{102},
047005 (2009).

\bibitem{umezawa} K. Umezawa {\it et al.}, Phys. Rev. Lett. {\bf 108}, 037002 (2012).

\bibitem{qureshi} N. Qureshi {\it et al.}, Phys. Rev. Lett. {\bf 108}, 117001 (2012).

\bibitem{aetaylor} A. E. Taylor {\it et al.}, Phys. Rev. B {\bf 83}, 220514 (R) (2011).

\bibitem{slli} S. L. Li \textit{et al.}, Phys. Rev. B \textbf{80}, 020504(R)
(2009).

\bibitem{drparker} D. R. Parker {\it et al.}, Phys. Rev. Lett. {\bf 104}, 057007 (2010).

\bibitem{xdzhou} X. D. Zhou {\it et al.}, arXiv:1204.4237.

\bibitem{clester10} C. Lester, J.-H. Chu, J. G. Analytis, T. G. Perring, I. R. Fisher, and S. M. Hayden, Phys. Rev. B {\bf 81}, 064505 (2010).

\bibitem{jtpark10} J. T. Park {\it et al.,}
Phys. Rev. B {\bf 82}, 134503 (2010).

\bibitem{huiqian} H. Q. Luo {\it et al.}, arXiv:1206.0653.

\bibitem{ykli} Y. K. Li {\it et al.}, New J. Phys. {\bf 12}, 083008 (2010).

\bibitem{mwang11} M. Wang {\it et al.}, Phys. Rev. B {\bf 83}, 220515 (R) (2011).

\bibitem{junli} J. Li {\it et al.}, Solid State Communications {\bf 152}, 671 (2012).

\bibitem{rmfernandez} R. M. Fernandes, M. G. Vavilov, and A. V. Chubukov, Phys. Rev. B {\bf 85}, 140512 (R) (2012).

\bibitem{tanatar} M. A. Tanatar {\it et al.}, Phys. Rev. B {\bf 85}, 014510 (2012).

\bibitem{Bang09} Y. Bang, H.Y. Choi, H. Won, Phys. Rev. B \textbf{79}, 054529 (2009).

\bibitem{putzke} C. Putzke {\it et al.}, Phys. Rev. Lett. {\bf 108}, 047002 (2012).

\bibitem{graser} S. Graser, A. F. Kemper, T. A. Maier, H.-P. Cheng, P. J.
Hirschfeld, and D. J. Scalapino, Phys. Rev. B {\bf 81}, 214503
(2010).

\bibitem{clzhang} C. L. Zhang {\it et al.}, Scienific Reports {\bf 1}, 115 (2011).

\bibitem{castellan} J.-P. Castellan {\it et al.}, Phys. Rev. Lett. {\bf 107}, 177003 (2011).

\end{thebibliography}

\end{document}